# Derivation of the Lorentz Transformations.

Rostislav Polishchuk
rpoluk@yahoo.co.uk

**Abstract:** The Lorentz Transformations are derived without any linearity assumptions and without assuming that y and z coordinates transform in a Galilean manner. Status of the invariance of the speed of light was reduced from a foundation of the Special Theory of Relativity to just a property which allows to determine a value of the physical constant. While high level of rigour is maintained, this paper should be accessible to a second year university physics student.

## I. Introduction

Over the last century a great number of derivations of the Lorentz transformations were published in textbooks, monographs and scientific journals. Apart from a few notable exceptions [1,2] all of them assume linearity of the transformation functions. Of those who do prove/discuss linearity, practically all assume that $y$ and $z$ coordinates (In standard configuration of coordinate systems) transform in a Galilean manner: i.e. $y' = y$ and $z' = z$.

In this paper I have attempted to derive the Lorentz transformations without making above mentioned assumptions. Status of the invariance of the speed of light was reduced from a foundation of the Special Theory of Relativity to just a property which allows to determine a value of the physical constant [3].

In this derivation I mainly used symmetry arguments and elementary algebra. In section on linearity partial derivatives theory is used.

## II. Definitions and Assumptions

Inertial coordinate system is defined as a coordinate system (CS) in which:
  a. Euclidean geometry applies
  b. Newton's first law holds

This definition immediately excludes the possibility of two coordinate systems whose origins accelerate relative to each other being inertial, as particle resting at the origin of one system will appear to accelerate in another.

The following assumptions are used:
1. Laws of physics are the same in all inertial coordinate systems. (Einstein's relativity principle) This implies isotropy of space and homogeneity of space and time.
2. Standard units of length and time are used in all inertial systems. (Relativity principle can be used to calibrate units of length and time in each inertial system)
3. All stationery observers equipped with standard clocks and located in the same inertial coordinate system will agree on temporal separation of events.
4. Transformation functions are differentiable at least twice with respect to coordinates and are continuous functions of the relative speed of the coordinate systems.
5. Any event that is "seen" in one inertial system is "seen" in all others. For example if observer in one system "sees" an explosion on a rocket then so do all other observers.
6. If (inertial) observer A "sees" that (inertial) observer B is moving radially away from him with speed $u$, then observer B "sees" observer A moving radially away from him with the same speed $u$. It might seem to the reader that this assumption can be derived from some symmetry arguments, however to apply symmetry arguments we would need to assume a theoretical possibility of existence of a third observer C, relative to which A and B are symmetrical.

Given one inertial coordinate system we can set-up second coordinate system whose origin A is at rest and axis are fixed (rather then rotating) as seen in first. As rules of Euclidean geometry apply in the first system they also apply in the second. Validity of Euclidean geometry implies that spatial separation of events is the same in both systems. By assumption (3) temporal separation also has to be the same, as observer located at the origin of second system is at rest relative to the first. Therefore in all inertial coordinate systems, whose origins are mutually at rest and axis are fixed, space-temporal separation of the events will be the same. Such systems are said to belong to the same frame and space-temporal separation of the events are property of the frame rather then of a particular coordinate system in it Any experiment can be described by referring to a particular frame but without a specific coordinate system in it. However to assign a particular values to coordinates we need a coordinate system.

Also it is clear that transformation functions must lead to one-to-one transformations otherwise single particle in one frame could appear as several (or have undetermined position) in another. Hence transformation functions have an inverse.

In this paper the following convention is used for representation of coordinates: $x_0 = t$, $x_1 = x$, $x_2 = y$, $x_3 = z$.

## III. Proof of linearity of coordinate transformations

### Part I

Consider particle P moving with a constant velocity **v** in an inertial coordinate system K. Let **r** be its position vector in K. Suppose every small time interval $\Delta\tau$, as measured in a rest frame of a particle, it flashes. Let $\Delta t$ be time interval between flashes as measured in K. Interval $\Delta t$ could be a function of **r**, **v** and $\Delta\tau$. (There are no other known parameters on which it could depend.)

However as observers of all coordinate systems of the same inertial frame agree on temporal separation of events it follows that $\Delta t$ could only be a function of speed of P and $\Delta\tau$. Otherwise we could set-up coordinate system K′ in the same frame as K and with axes parallel to corresponding axes of K but with a different origin and they would not agree on temporal separation of events (flashes); or we could set-up K′ in such a way that origins of both systems would coincide but with axes of K′ not-parallel to axis of K, and temporal separation of events would be different in these two systems.

It is clear that $\Delta t$ is directly proportional to $\Delta\tau$. (For example if we decide to measure time separation of every other flash, effectively doubling $\Delta\tau$ we will also double $\Delta t$.) Hence $\Delta t = f(\mathbf{v})\Delta\tau$. As transformation functions are one-to-one $f(v)$ is finite and is not equal to zero. By making $\Delta\tau$ infinitesimally small $dt = f(\mathbf{v})d\tau$ and $dt/d\tau = f(v)$.

Hence $\dfrac{dx_i}{d\tau} = \dfrac{dx_i}{dt}\dfrac{dt}{d\tau} = \dfrac{dx_i}{dt}f(v)$,

where $i = 0, 1, 2, 3$. As P moves with constant velocity $dx_i/dt$ and $f(\mathbf{v})$ are constant, and hence $dx_i/d\tau$ is constant.

Therefore for any particle moving with a constant velocity in an inertial frame $dx_i/d\tau$ are constant and $d^2x_i/d\tau^2$ are equal to zero, where $dx_i$ are separations of two events occurring on that particle as seen in the inertial frame and $d\tau$ is proper time interval between them.

## Part II
Consider inertial coordinate system K and inertial coordinate system K′ whose origin is moving with velocity $u$ relative to K. Suppose particle P moves with constant velocity $v$ in K. Then it moves with constant velocity, say $v'$ in K′. Using results from Part I for events occurring at P: $d^2x_i/d\tau^2$ and $d^2x'_i/d\tau^2$ are equal to zero. But:

$\dfrac{dx'_i}{d\tau} = \sum_{n=0}^{3}\dfrac{\partial x'_i}{\partial x_n}\dfrac{dx_n}{d\tau}$, and

$\dfrac{d^2x'_i}{d\tau^2} = \dfrac{d}{d\tau}\left(\dfrac{dx'_i}{d\tau}\right) = \dfrac{d}{d\tau}\left(\sum_{n=0}^{3}\dfrac{\partial x'_i}{\partial x_n}\dfrac{dx_n}{d\tau}\right) =$

$\sum_{n=0}^{3}\dfrac{d}{d\tau}\left(\dfrac{\partial x'_i}{\partial x_n}\right)\dfrac{dx_n}{d\tau} + \sum_{n=0}^{3}\dfrac{\partial x'_i}{\partial x_n}\dfrac{d^2x_n}{d\tau^2} =$

$\sum_{\substack{n=0\\m=0}}^{3}\dfrac{\partial^2 x'_i}{\partial x_n \partial x_m}\dfrac{dx_n}{d\tau}\dfrac{dx_m}{d\tau} + \sum_{n=0}^{3}\dfrac{\partial x'_i}{\partial x_n}\dfrac{d^2x_n}{d\tau^2}$.

Hence:

$\sum_{\substack{n=0\\m=0}}^{3}\dfrac{\partial^2 x'_i}{\partial x_n \partial x_m}\dfrac{dx_n}{d\tau}\dfrac{dx_m}{d\tau} = 0.$

As $dt/d\tau = f(v)$ and $v$ are constant:

$\sum_{\substack{n=0\\m=0}}^{3}\dfrac{\partial^2 x'_i}{\partial x_n \partial x_m}\dfrac{dx_n}{dt}\dfrac{dx_m}{dt} = 0.$

Last equation must be true at any point in space-time as we can imagine that it lies on a path of some particle.

As the components of the velocity (with exception of $dx_0/dt$ which is always equal to 1, but this does not invalidate the argument) of P are arbitrary it follows that at any point $\dfrac{\partial^2 x'_i}{\partial x_n \partial x_m} = 0$ and transformation functions must be linear functions of the coordinates i.e. be of the form: $x'_i = a_{ij}x_j + b_i$ (summation convention), where $a_{ij}$ and $b_i$ are constants for a particular configuration of the inertial coordinate systems. Inverse transformations must of the same form: $x_i = a'_{ij}x'_j + b'_i$.

## IV. Configuration of Coordinate Systems
Rather then trying to derive the most general transformations between two inertial coordinate systems we will consider a special case. However as transformations between coordinate systems of the same frame is the same as in classical kinematics, it is straightforward to derive most general transformations from the special.

Suppose we have an inertial coordinate systems K. Let origin of inertial coordinate system K′ move with speed $u$ along the positive direction of ox axis as seen in K. Choose ox′ of K′ in such a way that in K′ (in K′ means as seen by observer in K′) origin of K moves along the negative direction of x′ axis. Set $t$ and $t'$ in such a way that $t = t' = 0$ when origins of K and K′ coincide.

We did not yet completely define mutual configuration of K and K′ as we did not specify mutual orientation of $y$ and $y'$. Choice of orientation of one of these axes is arbitrary, however orientation of one of them defines orientation of another. I will return to this point later in this part of the paper.

All constants ($b_i$ and $b'_i$) in the transformation functions must be equal to zero otherwise origins will not coincide when $t = 0$ or $t' = 0$.

Coordinates $z$ and $y$ can not depend on $t'$ and coordinates $z'$ and $y'$ can not depend on $t$. Otherwise origin of one system would have a component of velocity perpendicular to $ox$ axis of another.

Consider two events on $y$ axis of K: one at $y = 1$ and another at $y = -1$. If they are simultaneous in K then they are also simultaneous in K′ and have their $x'$ coordinates are equal. Otherwise their temporal order or their $x'$ coordinates would be changed by reversal of $y$ axis (of K). As choice of $x'$ axis in K′ is not affected by orientation of $y$ axis, this is not allowed. Hence: coordinates $x'$ and $t'$ do not depend on $y$. Similarly: coordinates $x'$ and $t'$ do not depend on $z$ and coordinates $x$ and $t$ do not depend on $y'$ and $z'$.

Consider an event on the $x$ axis of K. As direction of $y'$ is arbitrary and is not determined by direction of $x$ axis it follows that the event can not have a non-zero $y'$ or

$z'$ component Hence $y'$ and $z'$ do not depend on $x$. Similarly, $y$ and $z$ do not depend on $x'$.

At this stage we have to define mutual configuration of K and K′ by specifying mutual orientation of $y'$ and $y$. We have shown that $y'$ and $z'$ do not depend on $x$ and $y$, and hence transformation functions are: $y' = a_{22}y + a_{23}z$ and $z' = a_{32}y + a_{33}z$. For any event, for which $z = 0$ transformation functions become: $y' = a_{22}y$ and $z' = a_{32}y$ and all such events lie on a straight line when projected onto $x'y'$ plane. Once $y$ axis is chosen we can choose $y'$ in such a way that it coincides with that line i.e. $a_{32} = 0$. Positive direction of $y'$ is chosen in such a way that $a_{22} > 0$. Then for any event:
$y' = a_{22}y + a_{23}z$ and
$z' = a_{33}z$.
It is clear from the last equation that for any event, for which $z' = 0$, $z$ is also equal to zero. Hence we can choose a direction of either of $y'$ and $y$ and then chose a direction of another in such a way that $z'$ (or $z$) is not a function of $y$ (or $y'$).

Consider two events for which: $x$ coordinates are equal, $t$ coordinates are equal, and one of them lies on $x$ axis. Then as $x'$ and $t'$ are not functions of $y$ and $z$, their $x'$ coordinates are equal and their $t'$ coordinates are equal. The one which lies on $x$-axis also lies on $x'$ axis as $y' = a_{22}y + a_{23}z$ and $z' = a_{33}z$ vanish for $y = z = 0$. Let the spatial separation of this events in K be $l$ and $l'$ in K′, then:
$l'^2 = y'^2 + z'^2 = a_{22}^2 y^2 + (a_{33}^2 + a_{23}^2)z^2 + 2a_{22}a_{23}yz$
where $y$ and $z$ are coordinates of the event which does not lie on $x$ axis. Unless $a_{23} = 0$; spatial separation of this events in K′ will depend on choice of $y$ in K (Consider case when $y = z = 1$, and rotate axis by 90° clockwise). Hence $l' = a_{22}^2 y^2 + a_{33}^2 z^2$. Also $a_{22}^2 = a_{33}^2$, otherwise $l'$ would be affected by choice of y axis in K again. Let $a_{22} = \mathbf{a}$, then: $l'^2 = \mathbf{a}^2 l^2$ and $l^2 = \frac{1}{\mathbf{a}^2}l'^2$. However if we reverse $x$, $x'$ $y$ and $y'$ (y axes are reversed to maintain handedness of the systems) roles of K and K′ are reversed, however spatial separation of this events should be left unchanged. Hence:
$\begin{cases} l'^2 = \mathbf{a}^2 l^2 \\ l^2 = \mathbf{a}^2 l'^2 \end{cases}$ and $\mathbf{a}^2 = 1$, $a_{22} = 1$.

As transformation functions are continuous functions of the velocity it follows that $a_{33} = 1$. Hence $y' = y$ and $z' = z$.

As origin of K′ is moving with speed $u$ along $x$ axis (of K) and $t=0$ when origins of K and K′ coincide it follows that for the origin of K′: $x = ut$,
$x' = a_{11}x + a_{10}t = a_{11}ut + a_{10}t = (a_{11}u + a_{10})t \equiv 0$.
Hence $a_{10} \equiv -a_{11}u$.

At this point, having reduced a number of coefficients, it makes sense to change notation. Let:
$a = a_{11}$, $b = a_{01}$, and $c = a_{00}$. Then:
$x' = a_{11}x + a_{10}t = a_{11}x - a_{11}ut = ax - aut$ and
$t' = a_{01}x + a_{00}t = bx + ct$.

Section Summary:
Choosing a particular configuration of coordinate systems, and applying symmetry arguments, allowed us to simplify transformation functions to:
$y' = y$,
$z' = z$,
$x' = ax - aut$,
$t' = bx + ct$,
where $a$, $b$, and $c$ are functions of $u$.

## V. Coefficients $b$ and $c$.

In this section coefficients $b$ and $c$ will be found in terms of $a$.

As reversing $x$ and $x'$ axis ($y$ and $y'$ would have to be reversed to maintain handedness of the coordinate systems, however in this section we concentrate on $x$ and $t$ coordinates.) will interchange roles of K and K′, it follows that:
$(-x) = a(-x') - aut'$ and $t = b(-x') + ct'$.
Hence:
$x = ax' + aut'$ and
$t = -bx' + ct'$

Clearly the last two equations are transformation functions from K′ to K. Transformation from K′ to K and then back must yield the original coordinates of an event. These requirements can be written in matrix form as:
$\begin{pmatrix} a & -au \\ b & c \end{pmatrix}\begin{pmatrix} a & au \\ -b & c \end{pmatrix} = \begin{pmatrix} a^2+abu & au(a-c) \\ b(a-c) & c^2+abu \end{pmatrix} = \begin{pmatrix} 1 & 0 \\ 0 & 1 \end{pmatrix}$
The above equation is satisfied only if $c = a$ and
$b = \frac{1-a^2}{au}$.
Hence, in matrix notation, transformation of $x$ and $t$ coordinates take the form:
$\begin{pmatrix} x' \\ t' \end{pmatrix} = \begin{pmatrix} a & -au \\ \frac{1-a^2}{au} & a \end{pmatrix}\begin{pmatrix} x \\ t \end{pmatrix}$, where $a$ is a function of $u$.

## VI. Coefficient $a$ as a function of $u$.
In this section we will find functional dependence of the coefficient $a$ on speed $u$.

Consider a third coordinate system K″, such that mutual configuration of K′ and K″ is the same as mutual configuration of K and K′ but with speed $v$ instead of $u$. We will now show that mutual configuration of K and K″ is the same as mutual configuration of K and K′ but with mutual speed, say, $w$ instead of $u$. This is essentially a proof of closure of the transformations.

Transformation functions from K′ to K″ are the same as transformation functions from K to K′ with $u=v$. Hence for any event $y″ = y′ = y$ and $z″ = z′ = z$. Therefore in K″ origin of K always lies on $x″$ axis and in K origin of K″ always lies on $x$-axis.

By construction, origins of K and K′ coincide when $t = t′ = 0$ and origins of K′ and K″ coincide when $t′ = t″ = 0$. Hence when $t = t″ = 0$ origins of K and K″ coincide.

Finally, in K origin of K″ moves along a positive direction of $x$-axis. Otherwise an observer in K would conclude that origin of K is in-between origins of K′ and K″, while observer in K′ would not find origin of K between origins of K′ and K″. (This would lead to contradiction as seen from the following example: Suppose a missile is located just tow the left of the origin of K (as seen in K) and a missile shield, which allows all objects apart from launched missiles to pass freely, is located at the origin of K. Then according to the observer A, located at the origin of K′, observer B, located at the origin of K″, could pick up the missile and shoot him down. However according to the observer in K, B would not be able to shoot down A as the shield would always be in a way of the missile.) Similarly, in K″ origin of K moves along the negative direction of the $x″$ axis and this completes the proof of closure.

As K and K″ are in the same configuration as K and K′ but with mutual speed $w$ instead of $u$, it follows that:

$$\begin{pmatrix} x″ \\ t″ \end{pmatrix} = \begin{pmatrix} a_w & -a_w w \\ \frac{1-a_w^2}{w} & a_w \end{pmatrix} \begin{pmatrix} x \\ t \end{pmatrix},$$ where $a_w$ stands for

$a(w)$ - the value of the coefficient $a$ when mutual speed of frames is equal to $w$.

However:

$$\begin{pmatrix} x″ \\ t″ \end{pmatrix} = \begin{pmatrix} a_v & -a_v v \\ \frac{1-a_w^2}{a_v v} & a_v \end{pmatrix} \begin{pmatrix} x′ \\ t′ \end{pmatrix} = \begin{pmatrix} a_v & -a_v v \\ \frac{1-a_w^2}{a_v v} & a_v \end{pmatrix} \begin{pmatrix} a_u & -a_u u \\ \frac{1-a_u^2}{a_u u} & a_u \end{pmatrix} \begin{pmatrix} x \\ t \end{pmatrix}$$

and hence:

$$\begin{pmatrix} a_w & -a_w w \\ \frac{1-a_w^2}{a_w w} & a_w \end{pmatrix} = \begin{pmatrix} a_v & -a_v v \\ \frac{1-a_v^2}{a_v v} & a_v \end{pmatrix} \begin{pmatrix} a_u & -a_u u \\ \frac{1-a_u^2}{a_u u} & a_u \end{pmatrix}.$$

As the upper-left element and the lower-right elements of the matrix to the right of the equality sign in the above equation are equal, it follows (from multiplication of matrices to the left of the equality sign) that:

$$a_v a_u - a_v v \left( \frac{1-a_u^2}{a_u u} \right) = -a_u u \left( \frac{1-a_v^2}{a_v v} \right) + a_v a_u \Rightarrow$$

$$a_v v \left( \frac{1-a_u^2}{a_u u} \right) = a_u u \left( \frac{1-a_v^2}{a_v v} \right) \Rightarrow$$

$$\frac{a_v^2 v^2}{1-a_v^2} = \frac{a_u^2 u^2}{1-a_u^2}.$$

As in the last equation left hand side depends only on $v$ and right hand side depends only on $u$, it follows that left hand side as well as right hand sides are constant. Let this constant be **m**, then:

$$\frac{a_u^2 u^2}{1-a_u^2} = \mathbf{m} \Rightarrow a_u^2 = \frac{1}{1+\frac{u^2}{\mathbf{m}}} \Rightarrow a_u = \sqrt{\frac{1}{1+\frac{u^2}{\mathbf{m}}}}.$$

Where positive square root was taken because for $u=0$ transformations must be identity transformations.

We have found the form of transformation functions between two frames in the standard configuration. The only remaining unknown is the value of the universal constant **m**. From the transformation functions we can make a number of conclusions, for example about possibility of time dilation. Then we could use time dilation experiments (involving decay of stationery and moving mesons) to measure the value of **m**. Another (but not the only) way to find a value of **m** is by deriving velocity addition formula and observing that if $\mathbf{m} = -c^2$ and an object moves with speed $c$ in one inertial frame, then it moves with the same speed in all others. This would enable us to identify $c$ as the speed of light in vacuum. But I stress once again that other experiments could be used to find the value of the constant.

Taking $\mathbf{m} = -c^2$, transformation function become:
$y′ = y,$
$z′ = z,$
$x′ = \frac{x-ut}{\sqrt{1-u^2/c^2}},$
$t′ = \frac{t-ux/c^2}{\sqrt{1-u^2/c^2}}.$

These transformation functions are known as the Lorentz transformations.